\documentclass[aip,apl, amssymb, amsmath, groupedaddress, reprint]{revtex4-1}

\usepackage{graphicx}
\usepackage[colorlinks=true,linkcolor=blue]{hyperref}
\usepackage{soul}
\usepackage[separate-uncertainty=true]{siunitx}
\usepackage{color}
\usepackage{wasysym}

\newcommand{\red}[1]{\textcolor{black}{#1}}
\usepackage{ulem}

\begin{document}

\title{A Cold Atom Radio-Frequency Magnetometer}

\author{Yuval Cohen}
\author{Krishna Jadeja}
\author{Sindi Sula}
\author{Michela Venturelli}
\author{Cameron Deans}
\author{Luca Marmugi}
\email{l.marmugi@ucl.ac.uk}
\author{\\Ferruccio Renzoni}
\affiliation{Department of Physics and Astronomy, University College London, Gower Street, London WC1E 6BT, United Kingdom}

\date{\today}

\begin{abstract}
We propose and demonstrate a radio-frequency atomic magnetometer with sub-Doppler laser cooled $\text{rubidium-}87$. With a simple and compact design, our system demonstrates a sensitivity \red{of \SI{330}{\pico\tesla\per\sqrt{\hertz}} in} an unshielded environment, thus matching or surpassing previously reported cold atoms designs. By merging the multiple uses and robustness of radio-frequency atomic magnetometers with the detailed control of laser cooling, our cold atom radio-frequency magnetometer has the potential to move applications of atomic magnetometry to high spatial resolution. Direct impact in metrology for applied sciences, materials characterization, and nanotechnology can be anticipated. 

\vskip 20pt
\begin{center}
This is a preprint version of the article appeared in Applied Physics Letters:\\
Y. Cohen, K. Jadeja, S. Sula, M. Venturelli, C. Deans, L. Marmugi, F. Renzoni,\\ Appl. Phys. Lett. {\bf 114}, 073505 (2019) DOI: \href{https://doi.org/10.1063/1.5084004}{10.1063/1.5084004}.
\end{center}
\end{abstract}


\maketitle

Magnetometry is at the core of many fundamental and applied sciences, with uses spanning from metrology to navigation. In this context, atomic magnetometers \cite{budker2007} (AMs) play an increasingly important role, mostly because of their sensitivity -- challenging superconducting quantum interference devices (SQUIDs) \cite{squidrsi2006}, without the need of cryogenics or, in some cases, magnetic shielding \cite{belfi2007,pustelny2008, bevilacqua2009,witold2012, subfemto, lucivero2014, bevilacqua2017}. This, paired with the relative simplicity of AMs \cite{schwindt2004,microam,serf2010}, triggered a rapid development of a wealth of instruments and applications with atomic samples at room temperature or above (thermal AMs)\cite{belfi2007, zulf,  edm1, darkmatter, nistcomm, nqr1, nqr2,  deans2016, witoldcui, jensen2018}. 

Although capable of great performance, thermal atomic magnetometers suffer from lack of control on atomic motion. Atomic spins are interrogated only as an ensemble spread across the interaction volume. This poses limitations in terms of the achievable spatial resolution, as well as on the proximity to the field source to be measured. More recently, Nitrogen-Vacancy (NV centers) magnetometers have improved the space and position performance \cite{nv, budker2016}, but the sensitivities of AMs remain unmatched.

On the contrary, cold or ultra-cold atoms systems are ideal candidates for high spatial resolution measurements. This is due to the fine control on the atomic degrees of freedom -- namely the suppression of thermal diffusion and collisions \cite{ptbec,kruger}. 


Observations of Faraday \red{rotation \cite{prafaraday, gawlik, new} and} demonstrations of magnetic field measurements  \cite{japanese, basonarxiv, kruger} with cold atoms have recently been achieved with a number of approaches. These were limited to \SI{}{\nano\tesla} sensitivities and above. Sensitivities in the \SI{10}{\pico\tesla} range were only obtained with a spinor Bose-Einstein condensate \cite{ptbec} and spin-squeezing \cite{spinsqueezing}. Similar sensitivities have also been predicted with spin-orbit-coupled atom interferometry \cite{durham2018}. These experiments require complex setups, and complicated protocols relying on the imaging of the atomic cloud. In addition, the measured sensitivities do not match those of thermal AMs, and/or the magnetic field measurement is only indirect. For these reasons, the practical use of cold AMs has been limited.

In this Letter, we propose, demonstrate, and characterize a cold atom radio-frequency (RF) AM \cite{savukov2005, witold2012} using $^{87}$Rb at  \SI{20}{\micro\kelvin}. After laser cooling, a stretched state sensitive to magnetic fields is created via optical pumping. An RF pulse excites atomic spin precession, which is probed by a low-intensity laser pulse. Faraday rotation of the probe beam's polarization plane allows read out of the spins motion and hence the measurement of a DC or AC field. Interrogation times exceeding \SI{20}{\milli\second} and a sensing volume between $\leq\SI{1}{\milli\metre\cubed}$ to $\SI{350}{\milli\metre\cubed}$ can be obtained. Compared to other cold atoms implementations, our system is more compact and robust. We achieve a \red{sensitivity\cite{savukov2005,witold2012} of $\delta B = \SI{330}{\pico\tesla\per\sqrt{\hertz}}$ in} an unshielded environment. This surpasses the sensitivities of cold magnetometers reported thus far in non-degenerate, non-squeezed gases. Our results pave the way to robust DC and AC magnetometry at the sub-mm scale, with potential for quick application in emerging fields such as high-resolution electromagnetic induction imaging \cite{deans2016, witoldcui} and characterization of surfaces and materials \cite{semiconductors, kruger}.

\begin{figure}[htbp]
\includegraphics[width=\linewidth]{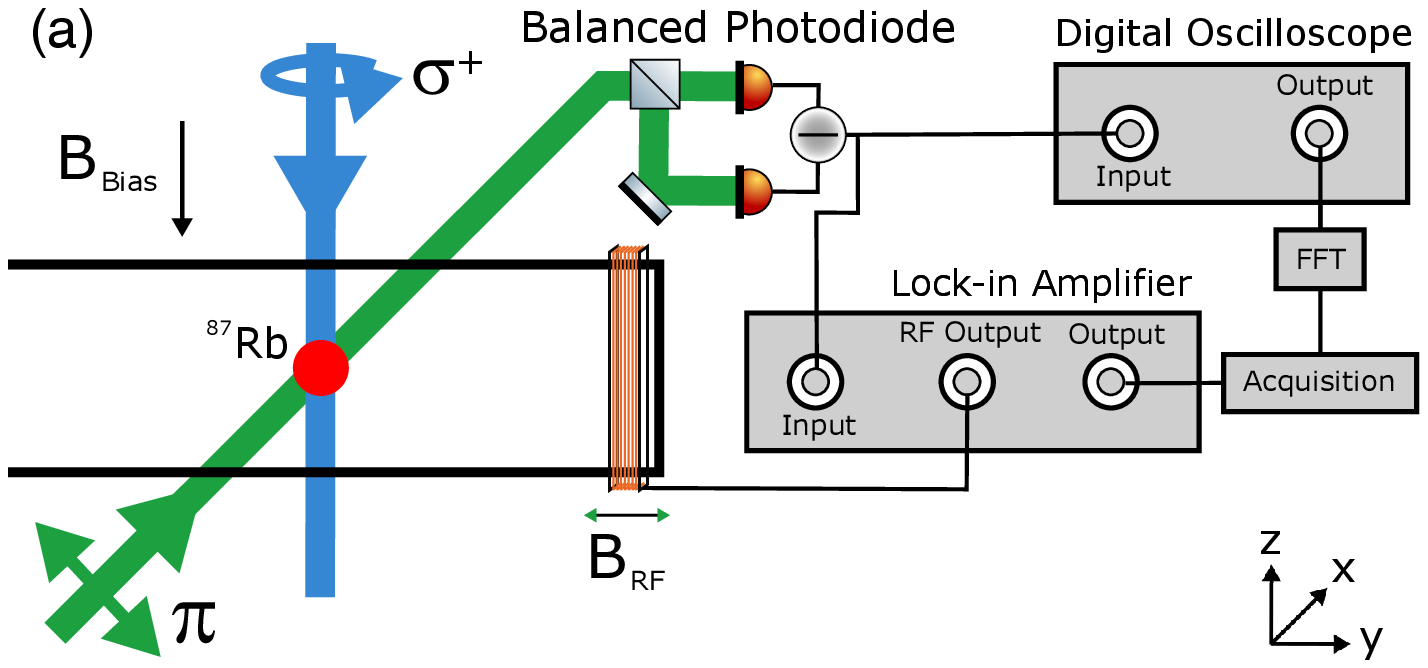}
\vskip 10pt
\includegraphics[width=\linewidth]{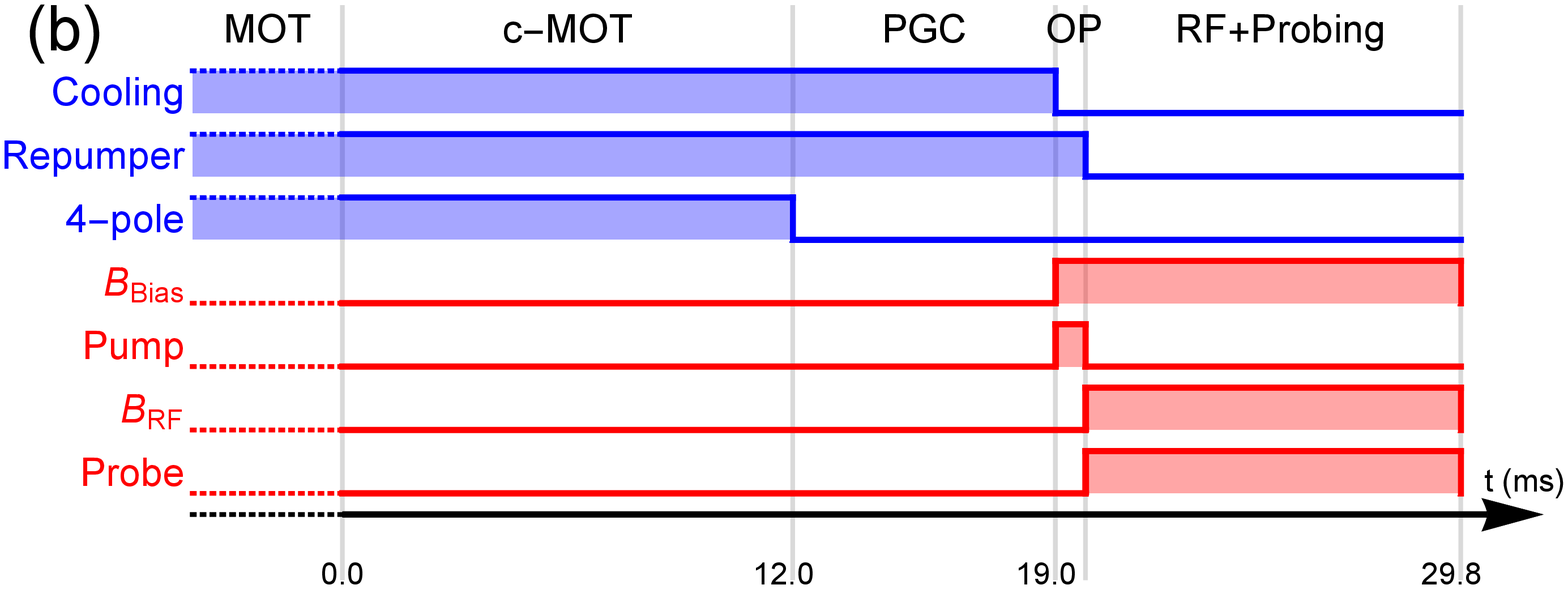}
\caption{(a) Sketch of the cold RF AM experimental setup. B$_{\text{Bias}}$ is the DC magnetic field for optical pumping. B$_{\text{RF}}$ is the AC magnetic field pulse driving the atomic precession. The blue line along the $-\hat{z}$ direction indicates the optical path of the $\sigma^{+}$ pump beam. The green line along $\hat{x}$ represents the $\pi$ polarized probe beam. (b) Overview of the experimental protocol. Details are in the main text. \label{fig:setup}}
\end{figure}

The experimental procedure can be divided in two distinct phases: preparation of cold atoms, and magnetometry. The design of the cold atoms section is derived from previous realizations \cite{rsibec}; a brief overview is provided here. An $^{87}$Rb 3D Magneto-Optical Trap (MOT) is loaded in a glass chamber of $\SI{30}{\milli\meter}\times\SI{30}{\milli\meter}\times\SI{100}{\milli\meter}$, from a Low-Velocity Intense Source (LVIS) \cite{lvis, rsibec}. A quadrupole magnetic field (4-pole in Fig.~\ref{fig:setup}) produces a linear gradient of \SI{32}{\text{G}\per\centi\meter} at the trapping region. Three pairs of compensation coils, in the ideal Helmholtz configuration, allow zeroing of the background field at the position of the atoms. The coils' supports are realized in PVC and 3D printed PLA to prevent spurious contributions from eddy currents. The cooling laser (supplied by a MOGLabs MSA003 system) is tuned $-1\Gamma$ to the red of the $^{87}$Rb $5s^{2}S_{1/2}F=2 \rightarrow 5p^{2}P_{3/2}F'=3$ transition, where $\Gamma=2\pi\cdot \SI{6.066}{\mega\hertz}$ is the natural linewidth. It has  an average intensity of \SI{7.5}{\milli\watt\per\centi\meter\squared} per MOT beam. The repumper laser (supplied by a MOGLabs CEL002 laser) is locked to the $^{87}$Rb $5s^{2}S_{1/2}F=1 \rightarrow 5p^{2}P_{3/2}F'=2$ transition, with an average intensity of  \SI{1.4}{\milli\watt\per\centi\meter\squared} per MOT beam. Computer-controlled acousto-optic modulators (AOMs)  allow fast switching of the different lasers. Real-time fluorescence imaging and destructive absorption imaging are used for monitoring and characterizing the atomic trap.  

The magnetometer phase relies on similar components to its thermal counterpart \cite{savukov2005,witold2012}. The core of this section of the experimental setup is sketched in Fig.~\ref{fig:setup}(a). 

A pair of circular Helmholtz coils (diameter \SI{80}{\milli\meter}) provides a homogeneous DC magnetic field along $-\hat{z}$ (B$_{\text{Bias}}$ in Fig.~\ref{fig:setup}(a)). This field sets the Zeeman splitting for optical pumping. Population transfer is produced by  a pulse of $\sigma^{+}$ polarized light, tuned to the $^{87}$Rb $5s^{2}S_{1/2}F=2 \rightarrow 5s^{2}P_{3/2}F'=3$ transition and propagating along $-\hat{z}$. The pump pulse -- controlled by a dedicated AOM -- has an intensity of \SI{60}{\micro\watt\per\centi\metre\squared}. A single RF coil (square cross-section, \SI{35}{\milli\metre} side, 20 turns, $\diameter=\SI{0.5}{\milli\meter}$ Cu wire, placed \SI{40}{\milli\meter} away from the center of the trapping region) is used to generate an AC magnetic field pulse along $\pm\hat{y}$ (B$_{\text{RF}}$ in Fig.~\ref{fig:setup}(a)). Atoms are probed by a $\pi$ polarized beam along $\hat{x}$, generated by a RadiantDyes NarrowDiode laser, tuned to $^{85}$Rb $5s^{2}S_{1/2}F=3 \rightarrow 5s^{2}P_{3/2}F'=3/4$ cross-over transition, thus  \SI{1.1}{\giga\hertz} to the blue of the pump laser. The typical intensity of the probe beam is \SI{257}{\micro\watt\per\centi\metre\squared}, actively stabilized by a dedicated AOM. The polarization of the probe beam is monitored by a polarimeter (formed by a polarizing beam splitter and a balanced photodiode -- See Fig.~\ref{fig:setup}(a)). Its output is fed to a digital oscilloscope (Tektronix DPO2014) and to a lock-in amplifier (AMETEK 7230 DSP), referenced to its internal local oscillator, which powers the RF coil. All instruments are connected to a computer-controlled acquisition system. In this way, the direct output of the cold RF AM, its Fast Fourier Transform (FFT), and the lock-in amplifier output are acquired, displayed, and stored.


The experimental sequence proceeds as sketched in Fig.~\ref{fig:setup}(b). Firstly, atoms are loaded for up to \SI{7}{\second} in the MOT. The atomic cloud is then compressed by ramping up the quadrupole field (up to a gradient of \SI{59}{G/\cm}) and detuning the cooling laser to -2$\Gamma$ in \SI{12}{\ms} (compressed MOT, c-MOT). \SI{7}{\milli\second} of polarization gradient cooling (PGC) bring the atoms' temperature down to T=$\SI{19\pm4}{\micro\kelvin}$, and to a density of \SI{e10}{\per\centi\meter\cubed}, while the cooling laser is further detuned to -4$\Gamma$, and its intensity decreased to around \SI{3}{\milli\watt\per\centi\meter\squared}. Up to $\SI{e8}{\text{atoms}}$ are trapped in the PGC phase. B$_{\text{Bias}}$ is then switched on and the optical pumping laser pulse irradiates the atomic sample for  \SI{0.8}{\milli\second}. This phase prepares the stretched state $|F=2, m_{F}=+2\rangle$ suitable for atomic magnetometry. Larmor precession is driven by a \SI{10}{\milli\second} RF pulse, triggered at the end of the optical pumping (OP) phase. At the same time, the probe pulse is fired, to monitor the Larmor precession. After the probing phase, the atomic cloud is dispersed or imaged with a resonant beam and a CCD camera for diagnostics. The repetition rate (approximately \SI{0.13}{\hertz}) is limited by the MOT loading time.

\begin{figure}[htbp]
\includegraphics[width=\linewidth]{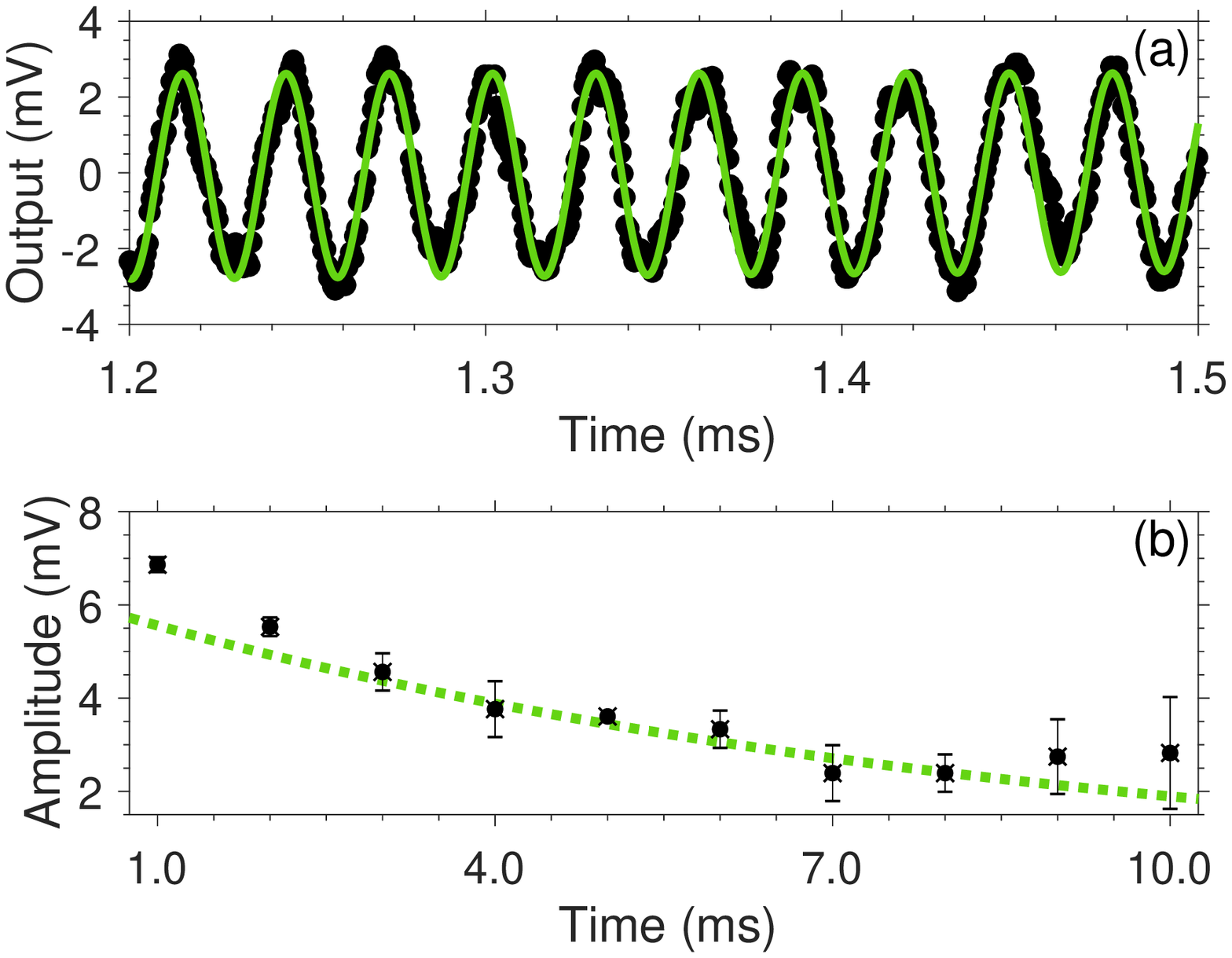}
\caption{Typical response of the cold RF AM. (a) Polarimeter output. The signal -- downsampled for clarity -- is fitted with an exponentially decaying sine function (in light green, overlapped to data points).  Data obtained at $\omega=\SI{34.5}{\kilo\hertz}$ (value from fit: $\omega_{\text{fit}}=\SI{34.49\pm 0.01}{\kilo\hertz}$) with RF pulse of \red{B$_{\text{RF}}=\SI{22}{\nano\tesla}$} (V$_{\text{rms}}=\SI{4.6}{\milli\volt}$). (b) Examples of the polarimeter output's amplitude versus interrogation time. The dashed line is the exponential fit, obtained from the total signal. \label{fig:oscillations}}
\end{figure}

The typical response of the cold RF AM is shown in Fig.~\ref{fig:oscillations}(a), where the polarimeter output during the probing phase is {\it continuously} acquired and plotted against time. 

Spins are coherently driven by the \SI{10}{\milli\second} RF pulse. Larmor precession around B$_{\text{Bias}}$ is produced, as a consequence of  the Zeeman transitions $|F=2, m_{F}\rangle \rightarrow  |F=2, m_{F\pm1}\rangle$ excited by B$_{\text{RF}}$. This process is mapped to the rotation of the polarization plane of the probe beam (Faraday rotation). The response is a sine wave oscillating at the Larmor frequency. At longer interrogation times, an exponential decay of the signal is observed, with a decay constant $\tau=\SI{8.4\pm0.3}{\milli\second}$ (Fig.~\ref{fig:oscillations}(b)).

The decay of the amplitude of the Faraday rotation signal is due to the progressive reduction of the atomic spins involved in the Larmor precession. We measured that the cold atoms' trap lifetime exceeds \SI{e2}{\milli\second}. Depolarizing collisions with background atoms can be excluded given the small residual pressure in the experimental chamber (P$\sim\SI{e-10}{\milli\bar}$). Therefore, the decay is attributed to the finite lifetime $\tau$ of the coherent precession, due to noise-induced spin dephasing. 

\begin{figure}[htbp]
\includegraphics[width=\linewidth]{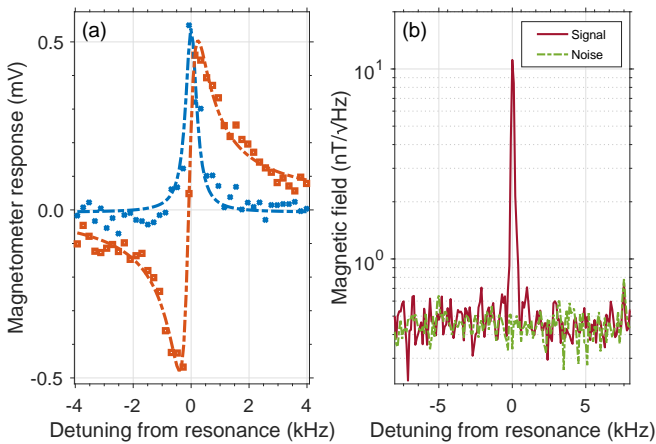}
\caption{Response of the cold RF AM. (a) Typical RF resonance, measured with the lock-in amplifier. Blue crosses: in-phase component (X). Red squares: quadrature component (Y). \red{The linewidth is \SI{230}{\hertz}}. Data were obtained with: N=\SI{1.1e8}{\text{atoms}}, and  \red{B$_{\text{RF}}=\SI{11}{\nano\tesla}$ }(V$_{\text{rms}}=\SI{2.3}{\milli\volt}$). (b) Typical FFT of the cold RF AM output, and technical noise level. Signal-to-noise ratio, SNR=33.}\label{fig:response}
\end{figure}

Figure~\ref{fig:response}(a) presents the resonance profile of the cold RF-AM at \SI{46.5}{\kilo\hertz}  (B$_{\text{Bias}}=\SI{66.4}{\milli\text{G}}$), measured with the lock-in amplifier. A Lorentzian response is observed in-phase with the RF driving. The quadrature component exhibits the typical dispersive behavior across the resonance.

Figure \ref{fig:response}(b) shows a typical FFT amplitude spectrum, as well as the total noise (measured in resonant conditions without the RF pulse). This is used to calculate the ratio between the signal and the noise (SNR) \cite{rsi2018}. We obtain an SNR of 33. After calibration of the SNR \red{with $\text{B}_{\text{RF}}=\SI{11}{\nano\tesla}$,} we measure a cold RF \red{AM sensitivity\cite{savukov2005, witold2012} of:}

\begin{equation}
\red{\delta B=\dfrac{B_{\text{RF}}}{\text{SNR}}=\SI{330}{\pico\tesla\per\sqrt{\hertz}}}~,\label{eqn:sensitivity}
\end{equation}

\noindent which surpasses recent realizations of cold atoms magnetometers by a factor $\geq$10. \red{Results from Eq.~\ref{eqn:sensitivity}  are consistent \cite{rsi2018} with those obtained with the alternative figure of merit\cite{pustelny2008,lucivero2014} $\delta B_{2}=\frac{\hbar}{g_{F} \mu_{B}}\frac{\Gamma}{\text{SNR}}$. Here, $\mu_{B}$ is the Bohr's magneton, $g_{F}$ the Land\'e factor, and $\Gamma$ the linewidth of the RF resonance.} This demonstrates the feasibility of a cold RF AM, with competitive performance, in a simpler setup with easier operation.



The width of the RF resonance (half-width at half-maximum, HWHM) is also investigated. Linewidths as low as HWHM=\SI{230}{\hertz} are obtained with V$_{\text{rms}}=\SI{2.3}{\milli\volt}$  supplied to the RF coil \red{($\text{B}_{\text{RF}}=\SI{11}{\nano\tesla}$)}. Power-induced broadening is observed with higher values of V$_{\text{rms}}$. In addition, splitting of the FFT spectrum is found for V$_{\text{rms}}>\SI{e3}{\milli\volt}$. Given its dependence on the RF power, this is interpreted as a Mollow triplet produced by dressing of the Zeeman states by the intense B$_{\text{RF}}$ \cite{valeriodressing, basondressing}. This effect substantially differs from the RF-induced broadening of the cold AM resonance measured with the lock-in amplifier.

\begin{figure}[htbp]
\includegraphics[width=\linewidth]{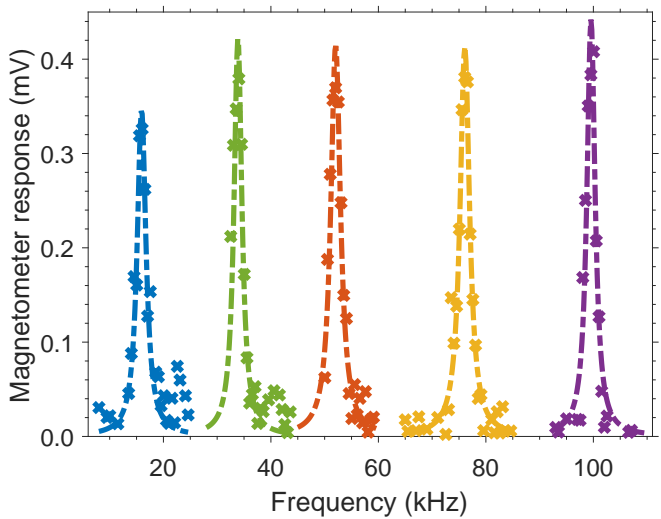}
\caption{Response at different frequencies (in-phase component X, measured via the lock-in amplifier), demonstrating the tunability of the cold RF AM. Dashed lines are the best Lorentzian fits of experimental data. Probed frequencies are $\omega_{\text{RF}}$=\SI{15.9}{}, \SI{33.9}{}, \SI{52.0}{}, \SI{76.1}{}, \SI{99.6}{\kilo\hertz} (\SI{22.7}{}, \SI{48.4}{}, \SI{74.3}{}, \SI{142.3}{\milli\text{G}}, respectively). \red{The width of the resonances -- around \SI{1}{\kilo\hertz} in this case -- does not exhibit relevant changes in the explored magnetic range.}}\label{fig:tunable}
\end{figure}

Our cold AM is also tunable over a wide band. In Fig.~\ref{fig:tunable}, operation between \SI{16}{\kilo\hertz} and \SI{100}{\kilo\hertz} is shown. A negligible broadening and a small variation in the signal's output are measured. This demonstrates consistent sub-\SI{}{\nano\tesla\per\sqrt{\hertz}} sensitivity across more than \SI{85}{\kilo\hertz}, limited on the upper bound by the lock-in amplifier bandwidth. 

These results show that the cold RF AM can operate also as a DC \red{scalar} magnetometer, measuring DC fields along the $\hat{z}$ direction. More specifically, in this configuration, changes in a DC magnetic field acting as -- or perturbing -- B$_{\text{Bias}}$ can be measured. Operation between \SI{22.7}{\milli\text{G}} and \SI{142.3}{\milli\text{G}} is demonstrated, \red{with no evidence of dependence of the HWHM -- and hence of the spins' dephasing rate -- on B$_{\text{Bias}}$.} In AC magnetometry, broad tunability is also an important resource for electromagnetic induction imaging\cite{opex2017,semiconductors}.

\begin{figure}[htbp]
\includegraphics[width=\linewidth]{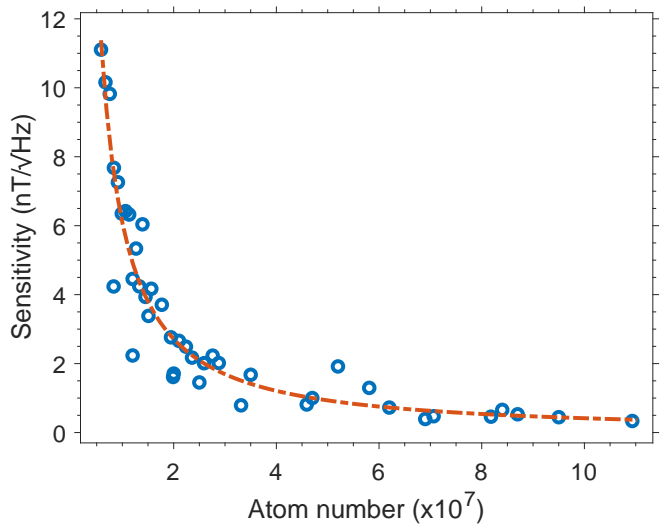}
\caption{Sensitivity \red{(Eq.~\ref{eqn:sensitivity})} of the cold RF AM versus number of atoms N in the polarization gradient cooling \red{(PGC)} phase. The dashed line is a fit with power N$^{-1.17}$.} \label{fig:natoms}
\end{figure}

The fine control over the atomic system allows additional parameters -- not possible in thermal AMs -- to be explored. An example is shown in Fig.~\ref{fig:natoms}, where the number of atoms after PGC -- hence immediately before the magnetometry phase -- is varied. The mean density of the core of the atomic cloud (around \SI{3e10}{\per\centi\meter\cubed}) remains approximately constant across the explored range.

The measured sensitivity (Eq.~\ref{eqn:sensitivity}) scales with the number of atoms $N$ according to the power law  $\delta B = aN^{b}$, \red{where $b=-1.17$, with a 95\% confidence interval $[-1.32, -1.01]_{0.95}$.} This suggests that the system's sensitivity is not limited by the spin-projection noise, as it is inconsistent with the scaling $\sim$N$^{-1/2}$ predicted and observed in thermal magnetometers \cite{savukov2005,spinnoise}. \red{Although various mechanisms may be at play \cite{new}, we attribute this to residual magnetic noise and gradients. Their impact could be reduced by using mu-metal shielding or active field stabilization \cite{rsi2018}.}


In conclusion, we have demonstrated a cold atom RF atomic magnetometer based on PGC of $^{87}$Rb, with sensitivity \red{of \SI{330}{\pico\tesla\per\sqrt{\hertz}} and} linewidth of \SI{230}{\hertz} in unshielded environments. Our approach relies on a simple and robust design. Pulsed operation with a repetition rate of \SI{0.13}{\hertz} was obtained. AC scalar magnetometry was demonstrated in the spectral region between \SI{16}{\kilo\hertz} and \SI{100}{\kilo\hertz}. Suitability for DC scalar magnetometry was also inferred from the presented measurements. The competitive performance allows one to envisage a future approach to atomic magnetometry, combining the ease of use of RF AMs with the high degree of control and spatial resolution provided by cold atoms. Short-term applications at the sub-\SI{}{\milli\metre} scale -- from passive magnetometry to active characterization of materials via electromagnetic induction imaging -- are anticipated.

\begin{acknowledgments}
This work was partially funded by the ERA-NET Cofund Transnational Call PhotonicSensing -- H2020 grant agreement No. 688735, with the project ``Magnetic Induction Tomography with Optical Sensors'' (MITOS)/Innovate UK Proj. No. 620129. Yuval Cohen is supported by the Engineering and Physical Sciences Research Council (EPSRC)  -- Grant EP/R512400/1. Krishna Jadeja is supported by the Engineering and Physical Sciences Research Council (EPSRC) -- Grant EP/N509395/1. Cameron Deans and Luca Marmugi acknowledge support from the UK Quantum Technology Hub in Sensing and Metrology, Engineering and Physical Sciences Research Council (EPSRC) (EP/M013294/1).
\end{acknowledgments}


\begin{thebibliography}{41}%
\makeatletter
\providecommand \@ifxundefined [1]{%
 \@ifx{#1\undefined}
}%
\providecommand \@ifnum [1]{%
 \ifnum #1\expandafter \@firstoftwo
 \else \expandafter \@secondoftwo
 \fi
}%
\providecommand \@ifx [1]{%
 \ifx #1\expandafter \@firstoftwo
 \else \expandafter \@secondoftwo
 \fi
}%
\providecommand \natexlab [1]{#1}%
\providecommand \enquote  [1]{``#1''}%
\providecommand \bibnamefont  [1]{#1}%
\providecommand \bibfnamefont [1]{#1}%
\providecommand \citenamefont [1]{#1}%
\providecommand \href@noop [0]{\@secondoftwo}%
\providecommand \href [0]{\begingroup \@sanitize@url \@href}%
\providecommand \@href[1]{\@@startlink{#1}\@@href}%
\providecommand \@@href[1]{\endgroup#1\@@endlink}%
\providecommand \@sanitize@url [0]{\catcode `\\12\catcode `\$12\catcode
  `\&12\catcode `\#12\catcode `\^12\catcode `\_12\catcode `\%12\relax}%
\providecommand \@@startlink[1]{}%
\providecommand \@@endlink[0]{}%
\providecommand \url  [0]{\begingroup\@sanitize@url \@url }%
\providecommand \@url [1]{\endgroup\@href {#1}{\urlprefix }}%
\providecommand \urlprefix  [0]{URL }%
\providecommand \Eprint [0]{\href }%
\providecommand \doibase [0]{http://dx.doi.org/}%
\providecommand \selectlanguage [0]{\@gobble}%
\providecommand \bibinfo  [0]{\@secondoftwo}%
\providecommand \bibfield  [0]{\@secondoftwo}%
\providecommand \translation [1]{[#1]}%
\providecommand \BibitemOpen [0]{}%
\providecommand \bibitemStop [0]{}%
\providecommand \bibitemNoStop [0]{.\EOS\space}%
\providecommand \EOS [0]{\spacefactor3000\relax}%
\providecommand \BibitemShut  [1]{\csname bibitem#1\endcsname}%
\let\auto@bib@innerbib\@empty
\bibitem [{\citenamefont {Budker}\ and\ \citenamefont
  {Romalis}(2007)}]{budker2007}%
  \BibitemOpen
  \bibfield  {author} {\bibinfo {author} {\bibfnamefont {D.}~\bibnamefont
  {Budker}}\ and\ \bibinfo {author} {\bibfnamefont {M.}~\bibnamefont
  {Romalis}},\ }\href {http://dx.doi.org/10.1038/nphys566} {\bibfield
  {journal} {\bibinfo  {journal} {Nature Physics}\ }\textbf {\bibinfo {volume}
  {3}},\ \bibinfo {pages} {227 EP } (\bibinfo {year} {2007})}\BibitemShut
  {NoStop}%
\bibitem [{\citenamefont {Fagaly}(2006)}]{squidrsi2006}%
  \BibitemOpen
  \bibfield  {author} {\bibinfo {author} {\bibfnamefont {R.~L.}\ \bibnamefont
  {Fagaly}},\ }\href {\doibase 10.1063/1.2354545} {\bibfield  {journal}
  {\bibinfo  {journal} {Review of Scientific Instruments}\ }\textbf {\bibinfo
  {volume} {77}},\ \bibinfo {pages} {101101} (\bibinfo {year} {2006})},\
  \Eprint {http://arxiv.org/abs/https://doi.org/10.1063/1.2354545}
  {https://doi.org/10.1063/1.2354545} \BibitemShut {NoStop}%
\bibitem [{\citenamefont {Belfi}\ \emph {et~al.}(2007)\citenamefont {Belfi},
  \citenamefont {Bevilacqua}, \citenamefont {Biancalana}, \citenamefont
  {Cartaleva}, \citenamefont {Dancheva},\ and\ \citenamefont
  {Moi}}]{belfi2007}%
  \BibitemOpen
  \bibfield  {author} {\bibinfo {author} {\bibfnamefont {J.}~\bibnamefont
  {Belfi}}, \bibinfo {author} {\bibfnamefont {G.}~\bibnamefont {Bevilacqua}},
  \bibinfo {author} {\bibfnamefont {V.}~\bibnamefont {Biancalana}}, \bibinfo
  {author} {\bibfnamefont {S.}~\bibnamefont {Cartaleva}}, \bibinfo {author}
  {\bibfnamefont {Y.}~\bibnamefont {Dancheva}}, \ and\ \bibinfo {author}
  {\bibfnamefont {L.}~\bibnamefont {Moi}},\ }\href {\doibase
  10.1364/JOSAB.24.002357} {\bibfield  {journal} {\bibinfo  {journal} {J. Opt.
  Soc. Am. B}\ }\textbf {\bibinfo {volume} {24}},\ \bibinfo {pages} {2357}
  (\bibinfo {year} {2007})}\BibitemShut {NoStop}%
\bibitem [{\citenamefont {Pustelny}\ \emph {et~al.}(2008)\citenamefont
  {Pustelny}, \citenamefont {Wojciechowski}, \citenamefont {Gring},
  \citenamefont {Kotyrba}, \citenamefont {Zachorowski},\ and\ \citenamefont
  {Gawlik}}]{pustelny2008}%
  \BibitemOpen
  \bibfield  {author} {\bibinfo {author} {\bibfnamefont {S.}~\bibnamefont
  {Pustelny}}, \bibinfo {author} {\bibfnamefont {A.}~\bibnamefont
  {Wojciechowski}}, \bibinfo {author} {\bibfnamefont {M.}~\bibnamefont
  {Gring}}, \bibinfo {author} {\bibfnamefont {M.}~\bibnamefont {Kotyrba}},
  \bibinfo {author} {\bibfnamefont {J.}~\bibnamefont {Zachorowski}}, \ and\
  \bibinfo {author} {\bibfnamefont {W.}~\bibnamefont {Gawlik}},\ }\href
  {\doibase 10.1063/1.2844494} {\bibfield  {journal} {\bibinfo  {journal}
  {Journal of Applied Physics}\ }\textbf {\bibinfo {volume} {103}},\ \bibinfo
  {pages} {063108} (\bibinfo {year} {2008})},\ \Eprint
  {http://arxiv.org/abs/https://doi.org/10.1063/1.2844494}
  {https://doi.org/10.1063/1.2844494} \BibitemShut {NoStop}%
\bibitem [{\citenamefont {Bevilacqua}\ \emph {et~al.}(2009)\citenamefont
  {Bevilacqua}, \citenamefont {Biancalana}, \citenamefont {Dancheva},\ and\
  \citenamefont {Moi}}]{bevilacqua2009}%
  \BibitemOpen
  \bibfield  {author} {\bibinfo {author} {\bibfnamefont {G.}~\bibnamefont
  {Bevilacqua}}, \bibinfo {author} {\bibfnamefont {V.}~\bibnamefont
  {Biancalana}}, \bibinfo {author} {\bibfnamefont {Y.}~\bibnamefont
  {Dancheva}}, \ and\ \bibinfo {author} {\bibfnamefont {L.}~\bibnamefont
  {Moi}},\ }\href {\doibase https://doi.org/10.1016/j.jmr.2009.09.013}
  {\bibfield  {journal} {\bibinfo  {journal} {Journal of Magnetic Resonance}\
  }\textbf {\bibinfo {volume} {201}},\ \bibinfo {pages} {222 } (\bibinfo {year}
  {2009})}\BibitemShut {NoStop}%
\bibitem [{\citenamefont {Chalupczak}\ \emph {et~al.}(2012)\citenamefont
  {Chalupczak}, \citenamefont {Godun}, \citenamefont {Pustelny},\ and\
  \citenamefont {Gawlik}}]{witold2012}%
  \BibitemOpen
  \bibfield  {author} {\bibinfo {author} {\bibfnamefont {W.}~\bibnamefont
  {Chalupczak}}, \bibinfo {author} {\bibfnamefont {R.~M.}\ \bibnamefont
  {Godun}}, \bibinfo {author} {\bibfnamefont {S.}~\bibnamefont {Pustelny}}, \
  and\ \bibinfo {author} {\bibfnamefont {W.}~\bibnamefont {Gawlik}},\ }\href
  {\doibase 10.1063/1.4729016} {\bibfield  {journal} {\bibinfo  {journal}
  {Applied Physics Letters}\ }\textbf {\bibinfo {volume} {100}},\ \bibinfo
  {pages} {242401} (\bibinfo {year} {2012})},\ \Eprint
  {http://arxiv.org/abs/https://doi.org/10.1063/1.4729016}
  {https://doi.org/10.1063/1.4729016} \BibitemShut {NoStop}%
\bibitem [{\citenamefont {Sheng}\ \emph {et~al.}(2013)\citenamefont {Sheng},
  \citenamefont {Li}, \citenamefont {Dural},\ and\ \citenamefont
  {Romalis}}]{subfemto}%
  \BibitemOpen
  \bibfield  {author} {\bibinfo {author} {\bibfnamefont {D.}~\bibnamefont
  {Sheng}}, \bibinfo {author} {\bibfnamefont {S.}~\bibnamefont {Li}}, \bibinfo
  {author} {\bibfnamefont {N.}~\bibnamefont {Dural}}, \ and\ \bibinfo {author}
  {\bibfnamefont {M.~V.}\ \bibnamefont {Romalis}},\ }\href {\doibase
  10.1103/PhysRevLett.110.160802} {\bibfield  {journal} {\bibinfo  {journal}
  {Phys. Rev. Lett.}\ }\textbf {\bibinfo {volume} {110}},\ \bibinfo {pages}
  {160802} (\bibinfo {year} {2013})}\BibitemShut {NoStop}%
\bibitem [{\citenamefont {Lucivero}\ \emph {et~al.}(2014)\citenamefont
  {Lucivero}, \citenamefont {Anielski}, \citenamefont {Gawlik},\ and\
  \citenamefont {Mitchell}}]{lucivero2014}%
  \BibitemOpen
  \bibfield  {author} {\bibinfo {author} {\bibfnamefont {V.~G.}\ \bibnamefont
  {Lucivero}}, \bibinfo {author} {\bibfnamefont {P.}~\bibnamefont {Anielski}},
  \bibinfo {author} {\bibfnamefont {W.}~\bibnamefont {Gawlik}}, \ and\ \bibinfo
  {author} {\bibfnamefont {M.~W.}\ \bibnamefont {Mitchell}},\ }\href {\doibase
  10.1063/1.4901588} {\bibfield  {journal} {\bibinfo  {journal} {Review of
  Scientific Instruments}\ }\textbf {\bibinfo {volume} {85}},\ \bibinfo {pages}
  {113108} (\bibinfo {year} {2014})},\ \Eprint
  {http://arxiv.org/abs/https://doi.org/10.1063/1.4901588}
  {https://doi.org/10.1063/1.4901588} \BibitemShut {NoStop}%
\bibitem [{\citenamefont {Bevilacqua}\ \emph {et~al.}(2017)\citenamefont
  {Bevilacqua}, \citenamefont {Biancalana}, \citenamefont {Dancheva},
  \citenamefont {Vigilante}, \citenamefont {Donati},\ and\ \citenamefont
  {Rossi}}]{bevilacqua2017}%
  \BibitemOpen
  \bibfield  {author} {\bibinfo {author} {\bibfnamefont {G.}~\bibnamefont
  {Bevilacqua}}, \bibinfo {author} {\bibfnamefont {V.}~\bibnamefont
  {Biancalana}}, \bibinfo {author} {\bibfnamefont {Y.}~\bibnamefont
  {Dancheva}}, \bibinfo {author} {\bibfnamefont {A.}~\bibnamefont {Vigilante}},
  \bibinfo {author} {\bibfnamefont {A.}~\bibnamefont {Donati}}, \ and\ \bibinfo
  {author} {\bibfnamefont {C.}~\bibnamefont {Rossi}},\ }\href {\doibase
  10.1021/acs.jpclett.7b02854} {\bibfield  {journal} {\bibinfo  {journal} {The
  Journal of Physical Chemistry Letters}\ }\textbf {\bibinfo {volume} {8}},\
  \bibinfo {pages} {6176} (\bibinfo {year} {2017})},\ \bibinfo {note} {pMID:
  29211488},\ \Eprint
  {http://arxiv.org/abs/https://doi.org/10.1021/acs.jpclett.7b02854}
  {https://doi.org/10.1021/acs.jpclett.7b02854} \BibitemShut {NoStop}%
\bibitem [{\citenamefont {Schwindt}\ \emph {et~al.}(2004)\citenamefont
  {Schwindt}, \citenamefont {Knappe}, \citenamefont {Shah}, \citenamefont
  {Hollberg}, \citenamefont {Kitching}, \citenamefont {Liew},\ and\
  \citenamefont {Moreland}}]{schwindt2004}%
  \BibitemOpen
  \bibfield  {author} {\bibinfo {author} {\bibfnamefont {P.~D.~D.}\
  \bibnamefont {Schwindt}}, \bibinfo {author} {\bibfnamefont {S.}~\bibnamefont
  {Knappe}}, \bibinfo {author} {\bibfnamefont {V.}~\bibnamefont {Shah}},
  \bibinfo {author} {\bibfnamefont {L.}~\bibnamefont {Hollberg}}, \bibinfo
  {author} {\bibfnamefont {J.}~\bibnamefont {Kitching}}, \bibinfo {author}
  {\bibfnamefont {L.-A.}\ \bibnamefont {Liew}}, \ and\ \bibinfo {author}
  {\bibfnamefont {J.}~\bibnamefont {Moreland}},\ }\href {\doibase
  10.1063/1.1839274} {\bibfield  {journal} {\bibinfo  {journal} {Applied
  Physics Letters}\ }\textbf {\bibinfo {volume} {85}},\ \bibinfo {pages} {6409}
  (\bibinfo {year} {2004})},\ \Eprint
  {http://arxiv.org/abs/https://doi.org/10.1063/1.1839274}
  {https://doi.org/10.1063/1.1839274} \BibitemShut {NoStop}%
\bibitem [{\citenamefont {Shah}\ \emph {et~al.}(2007)\citenamefont {Shah},
  \citenamefont {Knappe}, \citenamefont {Schwindt},\ and\ \citenamefont
  {Kitching}}]{microam}%
  \BibitemOpen
  \bibfield  {author} {\bibinfo {author} {\bibfnamefont {V.}~\bibnamefont
  {Shah}}, \bibinfo {author} {\bibfnamefont {S.}~\bibnamefont {Knappe}},
  \bibinfo {author} {\bibfnamefont {P.~D.~D.}\ \bibnamefont {Schwindt}}, \ and\
  \bibinfo {author} {\bibfnamefont {J.}~\bibnamefont {Kitching}},\ }\href
  {https://doi.org/10.1038/nphoton.2007.201} {\bibfield  {journal} {\bibinfo
  {journal} {Nature Photonics}\ }\textbf {\bibinfo {volume} {1}},\ \bibinfo
  {pages} {649 EP } (\bibinfo {year} {2007})}\BibitemShut {NoStop}%
\bibitem [{\citenamefont {Griffith}, \citenamefont {Knappe},\ and\
  \citenamefont {Kitching}(2010)}]{serf2010}%
  \BibitemOpen
  \bibfield  {author} {\bibinfo {author} {\bibfnamefont {W.~C.}\ \bibnamefont
  {Griffith}}, \bibinfo {author} {\bibfnamefont {S.}~\bibnamefont {Knappe}}, \
  and\ \bibinfo {author} {\bibfnamefont {J.}~\bibnamefont {Kitching}},\ }\href
  {\doibase 10.1364/OE.18.027167} {\bibfield  {journal} {\bibinfo  {journal}
  {Opt. Express}\ }\textbf {\bibinfo {volume} {18}},\ \bibinfo {pages} {27167}
  (\bibinfo {year} {2010})}\BibitemShut {NoStop}%
\bibitem [{\citenamefont {Ledbetter}\ \emph {et~al.}(2011)\citenamefont
  {Ledbetter}, \citenamefont {Theis}, \citenamefont {Blanchard}, \citenamefont
  {Ring}, \citenamefont {Ganssle}, \citenamefont {Appelt}, \citenamefont
  {Bl\"umich}, \citenamefont {Pines},\ and\ \citenamefont {Budker}}]{zulf}%
  \BibitemOpen
  \bibfield  {author} {\bibinfo {author} {\bibfnamefont {M.~P.}\ \bibnamefont
  {Ledbetter}}, \bibinfo {author} {\bibfnamefont {T.}~\bibnamefont {Theis}},
  \bibinfo {author} {\bibfnamefont {J.~W.}\ \bibnamefont {Blanchard}}, \bibinfo
  {author} {\bibfnamefont {H.}~\bibnamefont {Ring}}, \bibinfo {author}
  {\bibfnamefont {P.}~\bibnamefont {Ganssle}}, \bibinfo {author} {\bibfnamefont
  {S.}~\bibnamefont {Appelt}}, \bibinfo {author} {\bibfnamefont
  {B.}~\bibnamefont {Bl\"umich}}, \bibinfo {author} {\bibfnamefont
  {A.}~\bibnamefont {Pines}}, \ and\ \bibinfo {author} {\bibfnamefont
  {D.}~\bibnamefont {Budker}},\ }\href {\doibase
  10.1103/PhysRevLett.107.107601} {\bibfield  {journal} {\bibinfo  {journal}
  {Phys. Rev. Lett.}\ }\textbf {\bibinfo {volume} {107}},\ \bibinfo {pages}
  {107601} (\bibinfo {year} {2011})}\BibitemShut {NoStop}%
\bibitem [{\citenamefont {Baker}\ \emph {et~al.}(2014)\citenamefont {Baker},
  \citenamefont {Chibane}, \citenamefont {Chouder}, \citenamefont {Geltenbort},
  \citenamefont {Green}, \citenamefont {Harris}, \citenamefont {Heckel},
  \citenamefont {Iaydjiev}, \citenamefont {Ivanov}, \citenamefont {Kilvington},
  \citenamefont {Lamoreaux}, \citenamefont {May}, \citenamefont {Pendlebury},
  \citenamefont {Richardson}, \citenamefont {Shiers}, \citenamefont {Smith},\
  and\ \citenamefont {van~der Grinten}}]{edm1}%
  \BibitemOpen
  \bibfield  {author} {\bibinfo {author} {\bibfnamefont {C.}~\bibnamefont
  {Baker}}, \bibinfo {author} {\bibfnamefont {Y.}~\bibnamefont {Chibane}},
  \bibinfo {author} {\bibfnamefont {M.}~\bibnamefont {Chouder}}, \bibinfo
  {author} {\bibfnamefont {P.}~\bibnamefont {Geltenbort}}, \bibinfo {author}
  {\bibfnamefont {K.}~\bibnamefont {Green}}, \bibinfo {author} {\bibfnamefont
  {P.}~\bibnamefont {Harris}}, \bibinfo {author} {\bibfnamefont
  {B.}~\bibnamefont {Heckel}}, \bibinfo {author} {\bibfnamefont
  {P.}~\bibnamefont {Iaydjiev}}, \bibinfo {author} {\bibfnamefont
  {S.}~\bibnamefont {Ivanov}}, \bibinfo {author} {\bibfnamefont
  {I.}~\bibnamefont {Kilvington}}, \bibinfo {author} {\bibfnamefont
  {S.}~\bibnamefont {Lamoreaux}}, \bibinfo {author} {\bibfnamefont
  {D.}~\bibnamefont {May}}, \bibinfo {author} {\bibfnamefont {J.}~\bibnamefont
  {Pendlebury}}, \bibinfo {author} {\bibfnamefont {J.}~\bibnamefont
  {Richardson}}, \bibinfo {author} {\bibfnamefont {D.}~\bibnamefont {Shiers}},
  \bibinfo {author} {\bibfnamefont {K.}~\bibnamefont {Smith}}, \ and\ \bibinfo
  {author} {\bibfnamefont {M.}~\bibnamefont {van~der Grinten}},\ }\href
  {\doibase https://doi.org/10.1016/j.nima.2013.10.005} {\bibfield  {journal}
  {\bibinfo  {journal} {Nuclear Instruments and Methods in Physics Research
  Section A: Accelerators, Spectrometers, Detectors and Associated Equipment}\
  }\textbf {\bibinfo {volume} {736}},\ \bibinfo {pages} {184 } (\bibinfo {year}
  {2014})}\BibitemShut {NoStop}%
\bibitem [{\citenamefont {Jackson~Kimball}\ \emph {et~al.}(2018)\citenamefont
  {Jackson~Kimball}, \citenamefont {Budker}, \citenamefont {Eby}, \citenamefont
  {Pospelov}, \citenamefont {Pustelny}, \citenamefont {Scholtes}, \citenamefont
  {Stadnik}, \citenamefont {Weis},\ and\ \citenamefont
  {Wickenbrock}}]{darkmatter}%
  \BibitemOpen
  \bibfield  {author} {\bibinfo {author} {\bibfnamefont {D.~F.}\ \bibnamefont
  {Jackson~Kimball}}, \bibinfo {author} {\bibfnamefont {D.}~\bibnamefont
  {Budker}}, \bibinfo {author} {\bibfnamefont {J.}~\bibnamefont {Eby}},
  \bibinfo {author} {\bibfnamefont {M.}~\bibnamefont {Pospelov}}, \bibinfo
  {author} {\bibfnamefont {S.}~\bibnamefont {Pustelny}}, \bibinfo {author}
  {\bibfnamefont {T.}~\bibnamefont {Scholtes}}, \bibinfo {author}
  {\bibfnamefont {Y.~V.}\ \bibnamefont {Stadnik}}, \bibinfo {author}
  {\bibfnamefont {A.}~\bibnamefont {Weis}}, \ and\ \bibinfo {author}
  {\bibfnamefont {A.}~\bibnamefont {Wickenbrock}},\ }\href {\doibase
  10.1103/PhysRevD.97.043002} {\bibfield  {journal} {\bibinfo  {journal} {Phys.
  Rev. D}\ }\textbf {\bibinfo {volume} {97}},\ \bibinfo {pages} {043002}
  (\bibinfo {year} {2018})}\BibitemShut {NoStop}%
\bibitem [{\citenamefont {Gerginov}, \citenamefont {da~Silva},\ and\
  \citenamefont {Howe}(2017)}]{nistcomm}%
  \BibitemOpen
  \bibfield  {author} {\bibinfo {author} {\bibfnamefont {V.}~\bibnamefont
  {Gerginov}}, \bibinfo {author} {\bibfnamefont {F.~C.~S.}\ \bibnamefont
  {da~Silva}}, \ and\ \bibinfo {author} {\bibfnamefont {D.}~\bibnamefont
  {Howe}},\ }\href {\doibase 10.1063/1.5003821} {\bibfield  {journal} {\bibinfo
   {journal} {Review of Scientific Instruments}\ }\textbf {\bibinfo {volume}
  {88}},\ \bibinfo {pages} {125005} (\bibinfo {year} {2017})},\ \Eprint
  {http://arxiv.org/abs/https://doi.org/10.1063/1.5003821}
  {https://doi.org/10.1063/1.5003821} \BibitemShut {NoStop}%
\bibitem [{\citenamefont {Lee}\ \emph {et~al.}(2006)\citenamefont {Lee},
  \citenamefont {Sauer}, \citenamefont {Seltzer}, \citenamefont {Alem},\ and\
  \citenamefont {Romalis}}]{nqr1}%
  \BibitemOpen
  \bibfield  {author} {\bibinfo {author} {\bibfnamefont {S.-K.}\ \bibnamefont
  {Lee}}, \bibinfo {author} {\bibfnamefont {K.~L.}\ \bibnamefont {Sauer}},
  \bibinfo {author} {\bibfnamefont {S.~J.}\ \bibnamefont {Seltzer}}, \bibinfo
  {author} {\bibfnamefont {O.}~\bibnamefont {Alem}}, \ and\ \bibinfo {author}
  {\bibfnamefont {M.~V.}\ \bibnamefont {Romalis}},\ }\href {\doibase
  10.1063/1.2390643} {\bibfield  {journal} {\bibinfo  {journal} {Applied
  Physics Letters}\ }\textbf {\bibinfo {volume} {89}},\ \bibinfo {pages}
  {214106} (\bibinfo {year} {2006})},\ \Eprint
  {http://arxiv.org/abs/https://doi.org/10.1063/1.2390643}
  {https://doi.org/10.1063/1.2390643} \BibitemShut {NoStop}%
\bibitem [{\citenamefont {Cooper}\ \emph {et~al.}(2016)\citenamefont {Cooper},
  \citenamefont {Prescott}, \citenamefont {Matz}, \citenamefont {Sauer},
  \citenamefont {Dural}, \citenamefont {Romalis}, \citenamefont {Foley},
  \citenamefont {Kornack}, \citenamefont {Monti},\ and\ \citenamefont
  {Okamitsu}}]{nqr2}%
  \BibitemOpen
  \bibfield  {author} {\bibinfo {author} {\bibfnamefont {R.~J.}\ \bibnamefont
  {Cooper}}, \bibinfo {author} {\bibfnamefont {D.~W.}\ \bibnamefont
  {Prescott}}, \bibinfo {author} {\bibfnamefont {P.}~\bibnamefont {Matz}},
  \bibinfo {author} {\bibfnamefont {K.~L.}\ \bibnamefont {Sauer}}, \bibinfo
  {author} {\bibfnamefont {N.}~\bibnamefont {Dural}}, \bibinfo {author}
  {\bibfnamefont {M.~V.}\ \bibnamefont {Romalis}}, \bibinfo {author}
  {\bibfnamefont {E.~L.}\ \bibnamefont {Foley}}, \bibinfo {author}
  {\bibfnamefont {T.~W.}\ \bibnamefont {Kornack}}, \bibinfo {author}
  {\bibfnamefont {M.}~\bibnamefont {Monti}}, \ and\ \bibinfo {author}
  {\bibfnamefont {J.}~\bibnamefont {Okamitsu}},\ }\href {\doibase
  10.1103/PhysRevApplied.6.064014} {\bibfield  {journal} {\bibinfo  {journal}
  {Phys. Rev. Applied}\ }\textbf {\bibinfo {volume} {6}},\ \bibinfo {pages}
  {064014} (\bibinfo {year} {2016})}\BibitemShut {NoStop}%
\bibitem [{\citenamefont {Deans}\ \emph {et~al.}(2016)\citenamefont {Deans},
  \citenamefont {Marmugi}, \citenamefont {Hussain},\ and\ \citenamefont
  {Renzoni}}]{deans2016}%
  \BibitemOpen
  \bibfield  {author} {\bibinfo {author} {\bibfnamefont {C.}~\bibnamefont
  {Deans}}, \bibinfo {author} {\bibfnamefont {L.}~\bibnamefont {Marmugi}},
  \bibinfo {author} {\bibfnamefont {S.}~\bibnamefont {Hussain}}, \ and\
  \bibinfo {author} {\bibfnamefont {F.}~\bibnamefont {Renzoni}},\ }\href
  {\doibase 10.1063/1.4943659} {\bibfield  {journal} {\bibinfo  {journal}
  {Applied Physics Letters}\ }\textbf {\bibinfo {volume} {108}},\ \bibinfo
  {pages} {103503} (\bibinfo {year} {2016})},\ \Eprint
  {http://arxiv.org/abs/https://doi.org/10.1063/1.4943659}
  {https://doi.org/10.1063/1.4943659} \BibitemShut {NoStop}%
\bibitem [{\citenamefont {Bevington}\ \emph {et~al.}(2018)\citenamefont
  {Bevington}, \citenamefont {Gartman}, \citenamefont {Chalupczak},
  \citenamefont {Deans}, \citenamefont {Marmugi},\ and\ \citenamefont
  {Renzoni}}]{witoldcui}%
  \BibitemOpen
  \bibfield  {author} {\bibinfo {author} {\bibfnamefont {P.}~\bibnamefont
  {Bevington}}, \bibinfo {author} {\bibfnamefont {R.}~\bibnamefont {Gartman}},
  \bibinfo {author} {\bibfnamefont {W.}~\bibnamefont {Chalupczak}}, \bibinfo
  {author} {\bibfnamefont {C.}~\bibnamefont {Deans}}, \bibinfo {author}
  {\bibfnamefont {L.}~\bibnamefont {Marmugi}}, \ and\ \bibinfo {author}
  {\bibfnamefont {F.}~\bibnamefont {Renzoni}},\ }\href {\doibase
  10.1063/1.5042033} {\bibfield  {journal} {\bibinfo  {journal} {Applied
  Physics Letters}\ }\textbf {\bibinfo {volume} {113}},\ \bibinfo {pages}
  {063503} (\bibinfo {year} {2018})},\ \Eprint
  {http://arxiv.org/abs/https://doi.org/10.1063/1.5042033}
  {https://doi.org/10.1063/1.5042033} \BibitemShut {NoStop}%
\bibitem [{\citenamefont {Jensen}\ \emph {et~al.}(2018)\citenamefont {Jensen},
  \citenamefont {Skarsfeldt}, \citenamefont {St{\ae}rkind}, \citenamefont
  {Arnbak}, \citenamefont {Balabas}, \citenamefont {Olesen}, \citenamefont
  {Bentzen},\ and\ \citenamefont {Polzik}}]{jensen2018}%
  \BibitemOpen
  \bibfield  {author} {\bibinfo {author} {\bibfnamefont {K.}~\bibnamefont
  {Jensen}}, \bibinfo {author} {\bibfnamefont {M.~A.}\ \bibnamefont
  {Skarsfeldt}}, \bibinfo {author} {\bibfnamefont {H.}~\bibnamefont
  {St{\ae}rkind}}, \bibinfo {author} {\bibfnamefont {J.}~\bibnamefont
  {Arnbak}}, \bibinfo {author} {\bibfnamefont {M.~V.}\ \bibnamefont {Balabas}},
  \bibinfo {author} {\bibfnamefont {S.-P.}\ \bibnamefont {Olesen}}, \bibinfo
  {author} {\bibfnamefont {B.~H.}\ \bibnamefont {Bentzen}}, \ and\ \bibinfo
  {author} {\bibfnamefont {E.~S.}\ \bibnamefont {Polzik}},\ }\href {\doibase
  10.1038/s41598-018-34535-z} {\bibfield  {journal} {\bibinfo  {journal}
  {Scientific Reports}\ }\textbf {\bibinfo {volume} {8}},\ \bibinfo {pages}
  {16218} (\bibinfo {year} {2018})}\BibitemShut {NoStop}%
\bibitem [{\citenamefont {Rondin}\ \emph {et~al.}(2014)\citenamefont {Rondin},
  \citenamefont {Tetienne}, \citenamefont {Hingant}, \citenamefont {Roch},
  \citenamefont {Maletinsky},\ and\ \citenamefont {Jacques}}]{nv}%
  \BibitemOpen
  \bibfield  {author} {\bibinfo {author} {\bibfnamefont {L.}~\bibnamefont
  {Rondin}}, \bibinfo {author} {\bibfnamefont {J.-P.}\ \bibnamefont
  {Tetienne}}, \bibinfo {author} {\bibfnamefont {T.}~\bibnamefont {Hingant}},
  \bibinfo {author} {\bibfnamefont {J.-F.}\ \bibnamefont {Roch}}, \bibinfo
  {author} {\bibfnamefont {P.}~\bibnamefont {Maletinsky}}, \ and\ \bibinfo
  {author} {\bibfnamefont {V.}~\bibnamefont {Jacques}},\ }\href
  {http://stacks.iop.org/0034-4885/77/i=5/a=056503} {\bibfield  {journal}
  {\bibinfo  {journal} {Reports on Progress in Physics}\ }\textbf {\bibinfo
  {volume} {77}},\ \bibinfo {pages} {056503} (\bibinfo {year}
  {2014})}\BibitemShut {NoStop}%
\bibitem [{\citenamefont {Wickenbrock}\ \emph {et~al.}(2016)\citenamefont
  {Wickenbrock}, \citenamefont {Zheng}, \citenamefont {Bougas}, \citenamefont
  {Leefer}, \citenamefont {Afach}, \citenamefont {Jarmola}, \citenamefont
  {Acosta},\ and\ \citenamefont {Budker}}]{budker2016}%
  \BibitemOpen
  \bibfield  {author} {\bibinfo {author} {\bibfnamefont {A.}~\bibnamefont
  {Wickenbrock}}, \bibinfo {author} {\bibfnamefont {H.}~\bibnamefont {Zheng}},
  \bibinfo {author} {\bibfnamefont {L.}~\bibnamefont {Bougas}}, \bibinfo
  {author} {\bibfnamefont {N.}~\bibnamefont {Leefer}}, \bibinfo {author}
  {\bibfnamefont {S.}~\bibnamefont {Afach}}, \bibinfo {author} {\bibfnamefont
  {A.}~\bibnamefont {Jarmola}}, \bibinfo {author} {\bibfnamefont {V.~M.}\
  \bibnamefont {Acosta}}, \ and\ \bibinfo {author} {\bibfnamefont
  {D.}~\bibnamefont {Budker}},\ }\href {\doibase 10.1063/1.4960171} {\bibfield
  {journal} {\bibinfo  {journal} {Applied Physics Letters}\ }\textbf {\bibinfo
  {volume} {109}},\ \bibinfo {pages} {053505} (\bibinfo {year} {2016})},\
  \Eprint {http://arxiv.org/abs/https://doi.org/10.1063/1.4960171}
  {https://doi.org/10.1063/1.4960171} \BibitemShut {NoStop}%
\bibitem [{\citenamefont {Vengalattore}\ \emph {et~al.}(2007)\citenamefont
  {Vengalattore}, \citenamefont {Higbie}, \citenamefont {Leslie}, \citenamefont
  {Guzman}, \citenamefont {Sadler},\ and\ \citenamefont
  {Stamper-Kurn}}]{ptbec}%
  \BibitemOpen
  \bibfield  {author} {\bibinfo {author} {\bibfnamefont {M.}~\bibnamefont
  {Vengalattore}}, \bibinfo {author} {\bibfnamefont {J.~M.}\ \bibnamefont
  {Higbie}}, \bibinfo {author} {\bibfnamefont {S.~R.}\ \bibnamefont {Leslie}},
  \bibinfo {author} {\bibfnamefont {J.}~\bibnamefont {Guzman}}, \bibinfo
  {author} {\bibfnamefont {L.~E.}\ \bibnamefont {Sadler}}, \ and\ \bibinfo
  {author} {\bibfnamefont {D.~M.}\ \bibnamefont {Stamper-Kurn}},\ }\href
  {\doibase 10.1103/PhysRevLett.98.200801} {\bibfield  {journal} {\bibinfo
  {journal} {Phys. Rev. Lett.}\ }\textbf {\bibinfo {volume} {98}},\ \bibinfo
  {pages} {200801} (\bibinfo {year} {2007})}\BibitemShut {NoStop}%
\bibitem [{\citenamefont {Wildermuth}\ \emph {et~al.}(2006)\citenamefont
  {Wildermuth}, \citenamefont {Hofferberth}, \citenamefont {Lesanovsky},
  \citenamefont {Groth}, \citenamefont {Kr\"uger}, \citenamefont
  {Schmiedmayer},\ and\ \citenamefont {Bar-Joseph}}]{kruger}%
  \BibitemOpen
  \bibfield  {author} {\bibinfo {author} {\bibfnamefont {S.}~\bibnamefont
  {Wildermuth}}, \bibinfo {author} {\bibfnamefont {S.}~\bibnamefont
  {Hofferberth}}, \bibinfo {author} {\bibfnamefont {I.}~\bibnamefont
  {Lesanovsky}}, \bibinfo {author} {\bibfnamefont {S.}~\bibnamefont {Groth}},
  \bibinfo {author} {\bibfnamefont {P.}~\bibnamefont {Kr\"uger}}, \bibinfo
  {author} {\bibfnamefont {J.}~\bibnamefont {Schmiedmayer}}, \ and\ \bibinfo
  {author} {\bibfnamefont {I.}~\bibnamefont {Bar-Joseph}},\ }\href {\doibase
  10.1063/1.2216932} {\bibfield  {journal} {\bibinfo  {journal} {Applied
  Physics Letters}\ }\textbf {\bibinfo {volume} {88}},\ \bibinfo {pages}
  {264103} (\bibinfo {year} {2006})},\ \Eprint
  {http://arxiv.org/abs/https://doi.org/10.1063/1.2216932}
  {https://doi.org/10.1063/1.2216932} \BibitemShut {NoStop}%
\bibitem [{\citenamefont {Labeyrie}, \citenamefont {Miniatura},\ and\
  \citenamefont {Kaiser}(2001)}]{prafaraday}%
  \BibitemOpen
  \bibfield  {author} {\bibinfo {author} {\bibfnamefont {G.}~\bibnamefont
  {Labeyrie}}, \bibinfo {author} {\bibfnamefont {C.}~\bibnamefont {Miniatura}},
  \ and\ \bibinfo {author} {\bibfnamefont {R.}~\bibnamefont {Kaiser}},\ }\href
  {\doibase 10.1103/PhysRevA.64.033402} {\bibfield  {journal} {\bibinfo
  {journal} {Phys. Rev. A}\ }\textbf {\bibinfo {volume} {64}},\ \bibinfo
  {pages} {033402} (\bibinfo {year} {2001})}\BibitemShut {NoStop}%
\bibitem [{\citenamefont {Wojciechowski}\ \emph {et~al.}(2010)\citenamefont
  {Wojciechowski}, \citenamefont {Corsini}, \citenamefont {Zachorowski},\ and\
  \citenamefont {Gawlik}}]{gawlik}%
  \BibitemOpen
  \bibfield  {author} {\bibinfo {author} {\bibfnamefont {A.}~\bibnamefont
  {Wojciechowski}}, \bibinfo {author} {\bibfnamefont {E.}~\bibnamefont
  {Corsini}}, \bibinfo {author} {\bibfnamefont {J.}~\bibnamefont
  {Zachorowski}}, \ and\ \bibinfo {author} {\bibfnamefont {W.}~\bibnamefont
  {Gawlik}},\ }\href {\doibase 10.1103/PhysRevA.81.053420} {\bibfield
  {journal} {\bibinfo  {journal} {Phys. Rev. A}\ }\textbf {\bibinfo {volume}
  {81}},\ \bibinfo {pages} {053420} (\bibinfo {year} {2010})}\BibitemShut
  {NoStop}%
\bibitem [{\citenamefont {Sycz}, \citenamefont {Wojciechowski},\ and\
  \citenamefont {Gawlik}(2018)}]{new}%
  \BibitemOpen
  \bibfield  {author} {\bibinfo {author} {\bibfnamefont {K.}~\bibnamefont
  {Sycz}}, \bibinfo {author} {\bibfnamefont {A.~M.}\ \bibnamefont
  {Wojciechowski}}, \ and\ \bibinfo {author} {\bibfnamefont {W.}~\bibnamefont
  {Gawlik}},\ }\href {\doibase 10.1038/s41598-018-20522-x} {\bibfield
  {journal} {\bibinfo  {journal} {Scientific Reports}\ }\textbf {\bibinfo
  {volume} {8}},\ \bibinfo {pages} {2805} (\bibinfo {year} {2018})}\BibitemShut
  {NoStop}%
\bibitem [{\citenamefont {Isayama}\ \emph {et~al.}(1999)\citenamefont
  {Isayama}, \citenamefont {Takahashi}, \citenamefont {Tanaka}, \citenamefont
  {Toyoda}, \citenamefont {Ishikawa},\ and\ \citenamefont
  {Yabuzaki}}]{japanese}%
  \BibitemOpen
  \bibfield  {author} {\bibinfo {author} {\bibfnamefont {T.}~\bibnamefont
  {Isayama}}, \bibinfo {author} {\bibfnamefont {Y.}~\bibnamefont {Takahashi}},
  \bibinfo {author} {\bibfnamefont {N.}~\bibnamefont {Tanaka}}, \bibinfo
  {author} {\bibfnamefont {K.}~\bibnamefont {Toyoda}}, \bibinfo {author}
  {\bibfnamefont {K.}~\bibnamefont {Ishikawa}}, \ and\ \bibinfo {author}
  {\bibfnamefont {T.}~\bibnamefont {Yabuzaki}},\ }\href {\doibase
  10.1103/PhysRevA.59.4836} {\bibfield  {journal} {\bibinfo  {journal} {Phys.
  Rev. A}\ }\textbf {\bibinfo {volume} {59}},\ \bibinfo {pages} {4836}
  (\bibinfo {year} {1999})}\BibitemShut {NoStop}%
\bibitem [{\citenamefont {El{\'\i}asson}\ \emph {et~al.}(2018)\citenamefont
  {El{\'\i}asson}, \citenamefont {Heck}, \citenamefont {Laustsen},
  \citenamefont {Napolitano}, \citenamefont {M{\"u}ller}, \citenamefont
  {Bason}, \citenamefont {Arlt},\ and\ \citenamefont {Sherson}}]{basonarxiv}%
  \BibitemOpen
  \bibfield  {author} {\bibinfo {author} {\bibfnamefont {O.}~\bibnamefont
  {El{\'\i}asson}}, \bibinfo {author} {\bibfnamefont {R.}~\bibnamefont {Heck}},
  \bibinfo {author} {\bibfnamefont {J.~S.}\ \bibnamefont {Laustsen}}, \bibinfo
  {author} {\bibfnamefont {M.}~\bibnamefont {Napolitano}}, \bibinfo {author}
  {\bibfnamefont {R.}~\bibnamefont {M{\"u}ller}}, \bibinfo {author}
  {\bibfnamefont {M.~G.}\ \bibnamefont {Bason}}, \bibinfo {author}
  {\bibfnamefont {J.~J.}\ \bibnamefont {Arlt}}, \ and\ \bibinfo {author}
  {\bibfnamefont {J.~F.}\ \bibnamefont {Sherson}},\ }\href@noop {} {\bibfield
  {journal} {\bibinfo  {journal} {arXiv preprint arXiv:1811.01798}\ } (\bibinfo
  {year} {2018})}\BibitemShut {NoStop}%
\bibitem [{\citenamefont {Sewell}\ \emph {et~al.}(2012)\citenamefont {Sewell},
  \citenamefont {Koschorreck}, \citenamefont {Napolitano}, \citenamefont
  {Dubost}, \citenamefont {Behbood},\ and\ \citenamefont
  {Mitchell}}]{spinsqueezing}%
  \BibitemOpen
  \bibfield  {author} {\bibinfo {author} {\bibfnamefont {R.~J.}\ \bibnamefont
  {Sewell}}, \bibinfo {author} {\bibfnamefont {M.}~\bibnamefont {Koschorreck}},
  \bibinfo {author} {\bibfnamefont {M.}~\bibnamefont {Napolitano}}, \bibinfo
  {author} {\bibfnamefont {B.}~\bibnamefont {Dubost}}, \bibinfo {author}
  {\bibfnamefont {N.}~\bibnamefont {Behbood}}, \ and\ \bibinfo {author}
  {\bibfnamefont {M.~W.}\ \bibnamefont {Mitchell}},\ }\href {\doibase
  10.1103/PhysRevLett.109.253605} {\bibfield  {journal} {\bibinfo  {journal}
  {Phys. Rev. Lett.}\ }\textbf {\bibinfo {volume} {109}},\ \bibinfo {pages}
  {253605} (\bibinfo {year} {2012})}\BibitemShut {NoStop}%
\bibitem [{\citenamefont {Helm}\ \emph {et~al.}(2018)\citenamefont {Helm},
  \citenamefont {Billam}, \citenamefont {Rakonjac}, \citenamefont {Cornish},\
  and\ \citenamefont {Gardiner}}]{durham2018}%
  \BibitemOpen
  \bibfield  {author} {\bibinfo {author} {\bibfnamefont {J.~L.}\ \bibnamefont
  {Helm}}, \bibinfo {author} {\bibfnamefont {T.~P.}\ \bibnamefont {Billam}},
  \bibinfo {author} {\bibfnamefont {A.}~\bibnamefont {Rakonjac}}, \bibinfo
  {author} {\bibfnamefont {S.~L.}\ \bibnamefont {Cornish}}, \ and\ \bibinfo
  {author} {\bibfnamefont {S.~A.}\ \bibnamefont {Gardiner}},\ }\href {\doibase
  10.1103/PhysRevLett.120.063201} {\bibfield  {journal} {\bibinfo  {journal}
  {Phys. Rev. Lett.}\ }\textbf {\bibinfo {volume} {120}},\ \bibinfo {pages}
  {063201} (\bibinfo {year} {2018})}\BibitemShut {NoStop}%
\bibitem [{\citenamefont {Savukov}\ \emph {et~al.}(2005)\citenamefont
  {Savukov}, \citenamefont {Seltzer}, \citenamefont {Romalis},\ and\
  \citenamefont {Sauer}}]{savukov2005}%
  \BibitemOpen
  \bibfield  {author} {\bibinfo {author} {\bibfnamefont {I.~M.}\ \bibnamefont
  {Savukov}}, \bibinfo {author} {\bibfnamefont {S.~J.}\ \bibnamefont
  {Seltzer}}, \bibinfo {author} {\bibfnamefont {M.~V.}\ \bibnamefont
  {Romalis}}, \ and\ \bibinfo {author} {\bibfnamefont {K.~L.}\ \bibnamefont
  {Sauer}},\ }\href {\doibase 10.1103/PhysRevLett.95.063004} {\bibfield
  {journal} {\bibinfo  {journal} {Phys. Rev. Lett.}\ }\textbf {\bibinfo
  {volume} {95}},\ \bibinfo {pages} {063004} (\bibinfo {year}
  {2005})}\BibitemShut {NoStop}%
\bibitem [{\citenamefont {Marmugi}, \citenamefont {Deans},\ and\ \citenamefont
  {Renzoni}(2018)}]{semiconductors}%
  \BibitemOpen
  \bibfield  {author} {\bibinfo {author} {\bibfnamefont {L.}~\bibnamefont
  {Marmugi}}, \bibinfo {author} {\bibfnamefont {C.}~\bibnamefont {Deans}}, \
  and\ \bibinfo {author} {\bibfnamefont {F.}~\bibnamefont {Renzoni}},\
  }\href@noop {} {\bibfield  {journal} {\bibinfo  {journal} {arXiv preprint
  arXiv:1805.05743}\ } (\bibinfo {year} {2018})}\BibitemShut {NoStop}%
\bibitem [{\citenamefont {Nolli}\ \emph {et~al.}(2016)\citenamefont {Nolli},
  \citenamefont {Venturelli}, \citenamefont {Marmugi}, \citenamefont
  {Wickenbrock},\ and\ \citenamefont {Renzoni}}]{rsibec}%
  \BibitemOpen
  \bibfield  {author} {\bibinfo {author} {\bibfnamefont {R.}~\bibnamefont
  {Nolli}}, \bibinfo {author} {\bibfnamefont {M.}~\bibnamefont {Venturelli}},
  \bibinfo {author} {\bibfnamefont {L.}~\bibnamefont {Marmugi}}, \bibinfo
  {author} {\bibfnamefont {A.}~\bibnamefont {Wickenbrock}}, \ and\ \bibinfo
  {author} {\bibfnamefont {F.}~\bibnamefont {Renzoni}},\ }\href {\doibase
  10.1063/1.4960395} {\bibfield  {journal} {\bibinfo  {journal} {Review of
  Scientific Instruments}\ }\textbf {\bibinfo {volume} {87}},\ \bibinfo {pages}
  {083102} (\bibinfo {year} {2016})},\ \Eprint
  {http://arxiv.org/abs/https://doi.org/10.1063/1.4960395}
  {https://doi.org/10.1063/1.4960395} \BibitemShut {NoStop}%
\bibitem [{\citenamefont {Lu}\ \emph {et~al.}(1996)\citenamefont {Lu},
  \citenamefont {Corwin}, \citenamefont {Renn}, \citenamefont {Anderson},
  \citenamefont {Cornell},\ and\ \citenamefont {Wieman}}]{lvis}%
  \BibitemOpen
  \bibfield  {author} {\bibinfo {author} {\bibfnamefont {Z.~T.}\ \bibnamefont
  {Lu}}, \bibinfo {author} {\bibfnamefont {K.~L.}\ \bibnamefont {Corwin}},
  \bibinfo {author} {\bibfnamefont {M.~J.}\ \bibnamefont {Renn}}, \bibinfo
  {author} {\bibfnamefont {M.~H.}\ \bibnamefont {Anderson}}, \bibinfo {author}
  {\bibfnamefont {E.~A.}\ \bibnamefont {Cornell}}, \ and\ \bibinfo {author}
  {\bibfnamefont {C.~E.}\ \bibnamefont {Wieman}},\ }\href {\doibase
  10.1103/PhysRevLett.77.3331} {\bibfield  {journal} {\bibinfo  {journal}
  {Phys. Rev. Lett.}\ }\textbf {\bibinfo {volume} {77}},\ \bibinfo {pages}
  {3331} (\bibinfo {year} {1996})}\BibitemShut {NoStop}%
\bibitem [{\citenamefont {Deans}, \citenamefont {Marmugi},\ and\ \citenamefont
  {Renzoni}(2018)}]{rsi2018}%
  \BibitemOpen
  \bibfield  {author} {\bibinfo {author} {\bibfnamefont {C.}~\bibnamefont
  {Deans}}, \bibinfo {author} {\bibfnamefont {L.}~\bibnamefont {Marmugi}}, \
  and\ \bibinfo {author} {\bibfnamefont {F.}~\bibnamefont {Renzoni}},\ }\href
  {\doibase 10.1063/1.5026769} {\bibfield  {journal} {\bibinfo  {journal}
  {Review of Scientific Instruments}\ }\textbf {\bibinfo {volume} {89}},\
  \bibinfo {pages} {083111} (\bibinfo {year} {2018})},\ \Eprint
  {http://arxiv.org/abs/https://doi.org/10.1063/1.5026769}
  {https://doi.org/10.1063/1.5026769} \BibitemShut {NoStop}%
\bibitem [{\citenamefont {Bevilacqua}\ \emph {et~al.}(2012)\citenamefont
  {Bevilacqua}, \citenamefont {Biancalana}, \citenamefont {Dancheva},\ and\
  \citenamefont {Moi}}]{valeriodressing}%
  \BibitemOpen
  \bibfield  {author} {\bibinfo {author} {\bibfnamefont {G.}~\bibnamefont
  {Bevilacqua}}, \bibinfo {author} {\bibfnamefont {V.}~\bibnamefont
  {Biancalana}}, \bibinfo {author} {\bibfnamefont {Y.}~\bibnamefont
  {Dancheva}}, \ and\ \bibinfo {author} {\bibfnamefont {L.}~\bibnamefont
  {Moi}},\ }\href {\doibase 10.1103/PhysRevA.85.042510} {\bibfield  {journal}
  {\bibinfo  {journal} {Phys. Rev. A}\ }\textbf {\bibinfo {volume} {85}},\
  \bibinfo {pages} {042510} (\bibinfo {year} {2012})}\BibitemShut {NoStop}%
\bibitem [{\citenamefont {Jammi}\ \emph {et~al.}(2018)\citenamefont {Jammi},
  \citenamefont {Pyragius}, \citenamefont {Bason}, \citenamefont {Florez},\
  and\ \citenamefont {Fernholz}}]{basondressing}%
  \BibitemOpen
  \bibfield  {author} {\bibinfo {author} {\bibfnamefont {S.}~\bibnamefont
  {Jammi}}, \bibinfo {author} {\bibfnamefont {T.}~\bibnamefont {Pyragius}},
  \bibinfo {author} {\bibfnamefont {M.~G.}\ \bibnamefont {Bason}}, \bibinfo
  {author} {\bibfnamefont {H.~M.}\ \bibnamefont {Florez}}, \ and\ \bibinfo
  {author} {\bibfnamefont {T.}~\bibnamefont {Fernholz}},\ }\href {\doibase
  10.1103/PhysRevA.97.043416} {\bibfield  {journal} {\bibinfo  {journal} {Phys.
  Rev. A}\ }\textbf {\bibinfo {volume} {97}},\ \bibinfo {pages} {043416}
  (\bibinfo {year} {2018})}\BibitemShut {NoStop}%
\bibitem [{\citenamefont {Deans}, \citenamefont {Marmugi},\ and\ \citenamefont
  {Renzoni}(2017)}]{opex2017}%
  \BibitemOpen
  \bibfield  {author} {\bibinfo {author} {\bibfnamefont {C.}~\bibnamefont
  {Deans}}, \bibinfo {author} {\bibfnamefont {L.}~\bibnamefont {Marmugi}}, \
  and\ \bibinfo {author} {\bibfnamefont {F.}~\bibnamefont {Renzoni}},\ }\href
  {\doibase 10.1364/OE.25.017911} {\bibfield  {journal} {\bibinfo  {journal}
  {Opt. Express}\ }\textbf {\bibinfo {volume} {25}},\ \bibinfo {pages} {17911}
  (\bibinfo {year} {2017})}\BibitemShut {NoStop}%
\bibitem [{\citenamefont {S\o{}rensen}, \citenamefont {Hald},\ and\
  \citenamefont {Polzik}(1998)}]{spinnoise}%
  \BibitemOpen
  \bibfield  {author} {\bibinfo {author} {\bibfnamefont {J.~L.}\ \bibnamefont
  {S\o{}rensen}}, \bibinfo {author} {\bibfnamefont {J.}~\bibnamefont {Hald}}, \
  and\ \bibinfo {author} {\bibfnamefont {E.~S.}\ \bibnamefont {Polzik}},\
  }\href {\doibase 10.1103/PhysRevLett.80.3487} {\bibfield  {journal} {\bibinfo
   {journal} {Phys. Rev. Lett.}\ }\textbf {\bibinfo {volume} {80}},\ \bibinfo
  {pages} {3487} (\bibinfo {year} {1998})}\BibitemShut {NoStop}%
\end{thebibliography}
\end{document}